\documentclass[letterpaper,twocolumn,10pt]{article}
\usepackage{usenix2020}
\usepackage{url}
\usepackage{multirow}
\usepackage{adjustbox}
\usepackage{listings}
\usepackage{xspace}
\usepackage{amsmath}
\usepackage{booktabs}
\usepackage{comment}
\usepackage{enumitem}
\usepackage{subcaption}
\usepackage[ruled, vlined, linesnumbered]{algorithm2e}
\usepackage[normalem]{ulem}

\usepackage{hyperref}

\hypersetup{
  colorlinks,
  linkcolor={red!50!black},
  citecolor={blue!50!black},
  urlcolor={blue!80!black}
}

\microtypecontext{spacing=nonfrench}

\newcommand{\squishlist}{
   \begin{list}{$\bullet$}
    { \setlength{\itemsep}{0pt}      \setlength{\parsep}{3pt}
      \setlength{\topsep}{3pt}       \setlength{\partopsep}{0pt}
      \setlength{\leftmargin}{1.0em} \setlength{\labelwidth}{1em}
      \setlength{\labelsep}{0.5em} } }
\newcommand{\squishend}{
    \end{list}  }
    
\newcommand{\tool}[0]{ConServe\xspace}

\definecolor{javared}{rgb}{0.6,0,0} %
\definecolor{javagreen}{rgb}{0.25,0.5,0.35} %
\definecolor{javapurple}{rgb}{0.5,0,0.35} %
\definecolor{javadocblue}{rgb}{0.25,0.35,0.75} %

\lstdefinestyle{mystyle}{
      language=C++,
        basicstyle=\tiny\ttfamily,
        keywordstyle=\color{javapurple}\bfseries,
        stringstyle=\color{javared},
        commentstyle=\color{javadocblue},
        morecomment=[s][\color{javadocblue}]{/**}{*/},
        numbers=left,
        breaklines=true,
        numberstyle=\tiny\color{black},
        stepnumber=1,
        numbersep=5pt,
        tabsize=2,
        showspaces=false,
        showstringspaces=false,
        morekeywords={foreach, uint64_t, uint16_t},
        classoffset=0,
        xleftmargin=1.8em,
        escapeinside={(*@}{@*)},
        moredelim=*[is][\color{red}]{[[[}{]]]},
        captionpos=b, 
}
\newcommand{\us}[0]{µs\xspace}

\newcommand{\upth}[0]{\textsuperscript{th}\xspace}

\newcommand{\codeIn}[1]{{\small\texttt{#1}}}

\newcommand{\mysection}[1]{\section{#1}}
\newcommand{\mysubsection}[1]{\subsection{#1}}

\newcommand{\MyPara}[1]{\vspace{.3em}\noindent\textbf{#1}~}

\newcommand{\captionfonts}{\small}

\makeatletter  %
\long\def\@makecaption#1#2{%
  \vskip\abovecaptionskip
  \sbox\@tempboxa{{\captionfonts #1: #2}}%
  \ifdim \wd\@tempboxa >\hsize
    {\captionfonts #1: #2\par}
  \else
    \hbox to\hsize{\hfil\box\@tempboxa\hfil}%
  \fi
  \vskip\belowcaptionskip}
\makeatother   %

\newcommand{\squishlistree}{
   \begin{list}{$\bullet$}
    { \setlength{\itemsep}{0pt}      \setlength{\parsep}{0pt}
      \setlength{\topsep}{3pt}       \setlength{\partopsep}{0pt}
      \setlength{\leftmargin}{1em} \setlength{\labelwidth}{1em}
      \setlength{\labelsep}{0.5em} } }

\newcommand{\squishlisttwo}{
   \begin{list}{$\bullet$}
    { \setlength{\itemsep}{0pt}    \setlength{\parsep}{0pt}
      \setlength{\topsep}{0pt}     \setlength{\partopsep}{0pt}
      \setlength{\leftmargin}{2em} \setlength{\labelwidth}{1.5em}   
      \setlength{\labelsep}{0.5em} } }

\newcommand{\eg}{\hbox{\emph{e.g.}}\xspace}
\newcommand{\ie}{\hbox{\emph{i.e.}}\xspace}

\makeatletter
\newcommand*{\circled}{\@ifstar\circledstar\circlednostar}
\makeatother
 
\newcommand*\circledstar[1]{%
   \tikz[baseline=(C.base)]
     \node[%
       fill=black!20,
       circle,
       minimum size=1em,
       text=black,
       font=\footnotesize,
       inner sep=0.3pt
     ](C) {#1};%
}
\newcommand*\circlednostar[1]{%
   \tikz[baseline=(C.base) - .6em]
     \node[%
       fill=black,
       text=white,
       circle,
       minimum size=.8em,
       font={\bf \footnotesize},
       inner sep=0.2pt
     ](C) {#1};%
}

\begin{document}

\title{\tool: Fine-Grained GPU Harvesting for LLM Online and Offline Co-Serving}

\author{
\rm{Yifan Qiao}$^{\dagger}$\;
Shu Anzai$^{\ddagger}$\;
Shan Yu$^{\ddagger}$\;
Haoran Ma$^{\ddagger}$\;
Shuo Yang$^{\dagger}$\;
Yang Wang$^{\sharp}$\;
Miryung Kim$^{\ddagger}$\;\\[0.3em]
\rm{Yongji Wu}$^{\dagger}$\;
Yang Zhou$^{\dagger\clubsuit}$\; 
Jiarong Xing$^{\dagger\diamond}$\;
Joseph E. Gonzalez$^{\dagger}$\;
Ion Stoica$^{\dagger}$\; 
Harry Xu$^{\ddagger}$\;
\\[1em]
\rm{\emph{$^\dagger$UC Berkeley}}\; 
\emph{$^\ddagger$UCLA}\;
\emph{$^\diamond$Rice University}\; 
\emph{$^\clubsuit$UC Davis}\;
\emph{$^{\sharp}$Intel}
}

\maketitle

\begin{abstract}

Large language model (LLM) serving demands low latency and high throughput, but high load variability makes it challenging to achieve high GPU utilization. In this paper, we identify a synergetic but overlooked opportunity to co-serve latency-critical online requests alongside \textit{latency-tolerant offline} tasks such as model benchmarking. While promising, existing serving systems fail to co-serve them efficiently, as their coarse-grained resource management at the request or iteration level cannot harvest millisecond-level GPU idle cycles without introducing interference that violates online latency objectives.

\textit{\tool} is a new LLM co-serving system that achieves high throughput and strong online latency guarantees by managing resources at finer granularities. \tool introduces three techniques: (1) a latency-aware token-level scheduler that precisely sizes offline batches and tokens to fit within online latency objectives; (2) sub-iteration, layer-wise preemption that allows offline tasks to yield to online load spikes; and (3) incremental KV cache management that enables preempting and resuming offline requests at near-zero cost.
Evaluations with Llama-3.1 and Qwen-2.5 models on real-world workloads show that \tool delivers an average of 2.2$\times$ higher throughput and reduces online serving tail latency by 2.9$\times$ on average compared to state-of-the-art systems.

\end{abstract}

\mysection{Introduction}

Large language models (LLMs) 
are reshaping a broad range of applications such as chatbots~\cite{chatgpt,claude,chatbotarena@arxiv24}, coding assistants~\cite{copilot,codellama,jain2023llmassisted}, and document summarization~\cite{Narayan2018DontGM,notebooklm}. 
To meet rising demand, many providers\cite{openai-streaming-api,anthropic-streaming-api} now offer LLM online inference services, accessible through their APIs or web UI.
However, hosting LLMs entails substantial infrastructure costs. Serving millions of user requests~\cite{chatgpt-users, chatgpt-stats} demands running numerous model replicas across thousands or more high-performance GPUs. For instance, DeepSeek deploys \textasciitilde 3000 H800 GPUs to serve its R1 model~\cite{deepseek-system-overview}.

To make effective use of these costly infrastructures, LLM service providers always strive to offer low-latency services while maximizing their GPU utilization. 
However, in practice, the utilization of inference clusters is often unexpectedly low due to the bursty and unpredictable nature of online LLM workloads. 
These workloads can vary sharply, not just over hours but also within seconds. For example, a recent study~\cite{burstgpt} found that load of LLM services can triple within \emph{a few seconds}. 
Such rapid fluctuations are difficult for auto-scaling to handle, so today's providers often reserve long-term GPU instances or deploy on-premise clusters provisioned for peak load~\cite{xia2025skylb}.
This, in turn, leads to substantial GPU underutilization during low-traffic periods.

A common practice in the CPU world to mitigate resource waste is to backfill leftover cycles with best-effort jobs~\cite{shenango@nsdi19,zygos@sosp17,shinjuku@nsdi19}. 
Similar ideas are being explored for GPUs, such as harvesting leftover inference cycles for training or fine-tuning jobs~\cite{antman@osdi20,spotserve@asplos24,oliaro2025flexllmcoservinglargelanguage}.
The central challenge in such resource harvesting lies in ensuring that original workloads remain unaffected. In GPU harvesting, this requires that whenever online inference demands GPU resources, they can be reclaimed and reallocated seamlessly.
However, prior approaches have focused on colocated tasks such as training or fine-tuning, which typically run as independent processes in separate containers or virtual machines~\cite{serverlessllm@osdi24,zhang2024fastlivemodelauto}. Repartitioning GPU resources between these jobs and inference tasks is heavyweight, often incurring task-switching overheads that take seconds to minutes to complete.

To make GPU harvesting practical, it is critical to find best-effort jobs suitable for co-locating.
In this work, we argue that a more synergistic yet overlooked opportunity lies in co-serving online, latency-sensitive requests with \textit{offline}, latency-tolerant batch inference. This option is increasingly viable as providers like OpenAI~\cite{openai-batch-api} and Anthropic~\cite{anthropic-batch-api} now offer batch APIs with a 24-hour turnaround at a lower price for tasks such as
data analytics~\cite{liu2024optimizing} and model testing/evaluation~\cite{liang2023holisticevaluationlanguagemodels}.
Offline batch inference offers two key advantages for GPU harvesting: (1) offline tasks share the same model and can be co-scheduled with online ones in the same inference engine, enabling seamless reuse of the loaded model without extra provisioning and costly process switching; and (2) offline requests can be freely preempted within the engine for fast accommodation of sudden online traffic spikes.

However, simply colocating online and offline requests is insufficient due to the well-known tension between low latency and high efficiency~\cite{parties@asplos19,shenango@nsdi19,shinjuku@nsdi19}. Existing serving systems manage requests and their states at a coarse granularity that fails to mitigate interference and meet online workload's tight SLOs. Specifically, our profiling reveals three critical problems spanning the entire serving lifecycle: scheduling, execution, and KV cache state management.
\setlist[itemize]{topsep=0.3em, itemsep=0.1em, leftmargin=*}
\setlist[enumerate]{topsep=0.3em, itemsep=0.1em, leftmargin=*}
\begin{itemize}
\item \textit{Static, chunk-level scheduling creates unpredictable latency}. 
For online serving on overprovisioned clusters, schedulers such as continuous batching~\cite{orca@osdi22} are designed to maximize GPU throughput. They greedily enlarge batches as requests arrive, implicitly deprioritizing tail latency. Chunked prefill~\cite{sarathi-serve@arxiv24} partially regulates latency by enforcing a static compute budget (\ie, a fixed-size chunk), but still fills that budget rather than actively controlling latency. 
Under co-serving, where resources are intentionally saturated and every batch can be filled, this static, chunk-level approach becomes untenable. The scheduler must instead track the performance impact of admitting each request and token and dynamically adjust the token budget to guarantee online SLOs.

\item \textit{Iteration-level preemption causes slow responsiveness}. To handle a sudden spike in online traffic, offline work must be preempted immediately. However, to maximize the GPU efficiency, current serving engines decouple CPU and GPU execution: the scheduler forwards a batch to a dedicated GPU executor, which then runs the entire iteration uninterruptibly as a single executable or a CUDA Graph. As a result, the scheduler can only regain control and preempt after the current iteration completes. 
For example, our profiling shows that a large offline batch can introduce 970ms of delay for a Llama-3.1 8B model and stall online serving response.
\item \textit{Request-level state management incurs prohibitive overhead}.
Since an LLM needs the entire past context to generate a new token, the request serves as the basic unit for execution and state management. However, while this granularity simplifies scheduling, it is too coarse-grained for KV cache management, which can grow to gigabytes of state per request. As a result, preempting a request forces a costly choice: either discard the entire KV cache and pay for expensive recomputation later, or stall the GPU to swap it back and forth to host memory, leading to a collapse in throughput.
\end{itemize}
As a result, when co-serving LLM requests, existing systems suffer from a 3$\times$ online latency increase and 4$\times$ offline throughput drop (\S\ref{sec:motivation}).

\MyPara{Insight.} 
This paper addresses a central question: \emph{Can idle GPU cycles and memory be harvested during online LLM serving while achieving both high throughput and strict SLO guarantees?} 
Our key insight is that resolving the tension between low latency and high efficiency requires a co-serving system to manage resources and orchestrate execution at a granularity finer than requests or iterations. By operating at this finer scale, the system can exploit scheduling and preemption opportunities at the millisecond level, enabling both efficient utilization and strong SLO adherence.

\MyPara{Our Approach.}
To realize this insight, we introduce \tool, the first LLM co-serving system that harvests idle resources with the responsiveness to adhere to strong online SLOs. \tool orchestrates co-serving at the sub-request and sub-iteration level through three core techniques.
\begin{itemize}
\item \textbf{SLO-aware token-level scheduling} (\S\ref{sec:design:scheduler}). 
To fix the issue of unpredictable latency caused by static, chunk-level scheduling, \tool introduces a dynamic, fine-grained scheduler. It uses an offline profiler to build a precise cost model that captures how batch latency is affected by both compute pressure (active tokens) and memory pressure (cumulative KV cache). At runtime, the scheduler queries this model to dynamically adjust the number of offline tokens it adds to a batch, ensuring it precisely fits within the available SLO slack while maximizing throughput.
\item \textbf{Layer-wise preemption} (\S\ref{sec:design:preemption}). 
To break the iteration-level boundary that causes slow responsiveness, \tool enables preemption within a single forward pass. By instrumenting models with safe preemption points between transformer layers, \tool allows offline requests to exit mid-iteration upon receiving a preemption signal from the scheduler. This reduces preemption latency from the hundreds of milliseconds of a full iteration to the single-digit milliseconds of one or several layers, enabling near-instantaneous reaction to traffic spikes.
\item \textbf{Incremental KV cache management} (\S\ref{sec:design:inc-checkpointing}). 
To eliminate the high cost of recomputing or swapping KV caches upon preemption, \tool leverages a key observation that the KV cache is append-only and immutable. Specifically, \tool decouples a request from its full state history by managing the state at the token level. Instead of swapping the entire KV cache of a request, \tool asynchronously checkpoints only the state of the single, newly-generated token after each iteration per request, eliminating overhead that would otherwise be proportional to the sequence's length. This incremental approach, combined with background prefetching for restoration, makes frequent, millisecond-scale preemption practical at near-zero cost and allows GPU memory to be safely saturated without risking performance.
\end{itemize}

\MyPara{Results.}
We evaluated \tool with Llama-3.1 8B~\cite{grattafiori2024llama3herdmodels}, Qwen-2.5 14B~\cite{qwen2.5techreport@arxiv}, and Llama-3.1 70B~\cite{grattafiori2024llama3herdmodels} across four real-world datasets and a range of synthetic workloads. Results show that \tool significantly improves GPU utilization while preserving low online latency and high offline throughput.
Compared to state-of-the-art serving systems with preemptive scheduling, \tool provides strong performance isolation for online inference and reduces P99 latency by an average of 2.9$\times$, while improving the offline throughput by an average of 2.2$\times$.

\mysection{Background}

\vspace{0.8em}
\mysubsection{Large Language Model Inference}

Modern LLMs process a sequence of input tokens (\ie, prompt) and generate output tokens through two sequential phases: \codeIn{prefill} and \codeIn{decode}.
An LLM inference starts with the \codeIn{prefill} phase, where the model processes the entire input sequence to generate the first output token. 
An LLM can process all input tokens in parallel, making the \codeIn{prefill} phase compute-bound. 
Following that, the \codeIn{decode} phase generates subsequent tokens in an auto-regressive manner. 
For each output token, its generation requires the computation of a series of key and value vectors, which remain unchanged throughout the inference process. Therefore, the inference engine~\cite{ott2019fairseq, fastertransformer,tensorrt-llm} caches them in GPU memory (\ie, KV cache) to avoid recomputation in each \codeIn{decode} step. 
Due to the auto-regressive generation process and the design of the KV cache, the \codeIn{decode} phase is less compute-intensive and typically bounded by the GPU memory bandwidth due to the need to load model weights.

To improve the compute intensity and hardware efficiency during LLM inference, a common practice is to batch multiple requests in a single inference run to reuse the model weights for computation and memory load~\cite{orca@osdi22,vllm@sosp23}.

\mysubsection{Online and Offline LLM Serving}
Broadly, LLM serving can be categorized into two types:
online serving, which generates real-time responses to user inputs, and offline serving, which processes inputs in batches and responds asynchronously. 

Online serving is primarily designed for latency-critical requests and is the de facto option offered by existing API providers~\cite{openai-streaming-api, anthropic-streaming-api}.
Unlike traditional cloud services, where processing times are often predictable, an LLM request generates a sequence of tokens with variable lengths. As a result, LLM serving latency is measured on a per-token basis. 
Specifically, an LLM's serving latency is defined by two key metrics: \emph{time to first token (TTFT)}, which is the duration of the \codeIn{prefill} phase, and \emph{time between tokens (TBT)}, which is the execution time of each step in the \codeIn{decode} phase\footnote{
Note that TBT is measured \textit{per token} and is stricter than time per output token (TPOT), which averages the TBT across all tokens \textit{per request}.
}.
To provide a low end-to-end response latency and smooth user experience, it is critical for an online LLM serving system to optimize both TTFT and TBT under stringent SLOs. For example, an online chatbot may set a 99\upth-percentile (P99) TTFT SLO of 1.5 seconds and a P99 TBT SLO of 100 milliseconds to respond faster than a typical human can read~\cite{llm-problem-latency@blog,llm-best-practice@blog}.

Offline serving offers a cost-efficient solution for latency-tolerant, best-effort requests. Users typically submit offline requests in batches, which are processed by the serving system in a best-effort manner to maximize hardware efficiency. Unlike online serving which emphasizes low latency, offline serving often involves processing large corpora or datasets, where the first-order performance metric is \emph{throughput} that measures how many tokens are generated per second.
Typical scenarios for offline serving include document summarization~\cite{jin2024comprehensivesurveyprocessorientedautomatic}, data wrangling~\cite{narayan2022foundationmodelswrangledata}, data analytics~\cite{liu2024optimizing}, LLM benchmarking and evaluation ~\cite{liang2023holisticevaluationlanguagemodels}, among other emerging applications.

\begin{figure}[t]
    \centering
    \small
    \begin{subfigure}[b]{0.9\linewidth}
        \centering
        \includegraphics[width=0.99\linewidth]{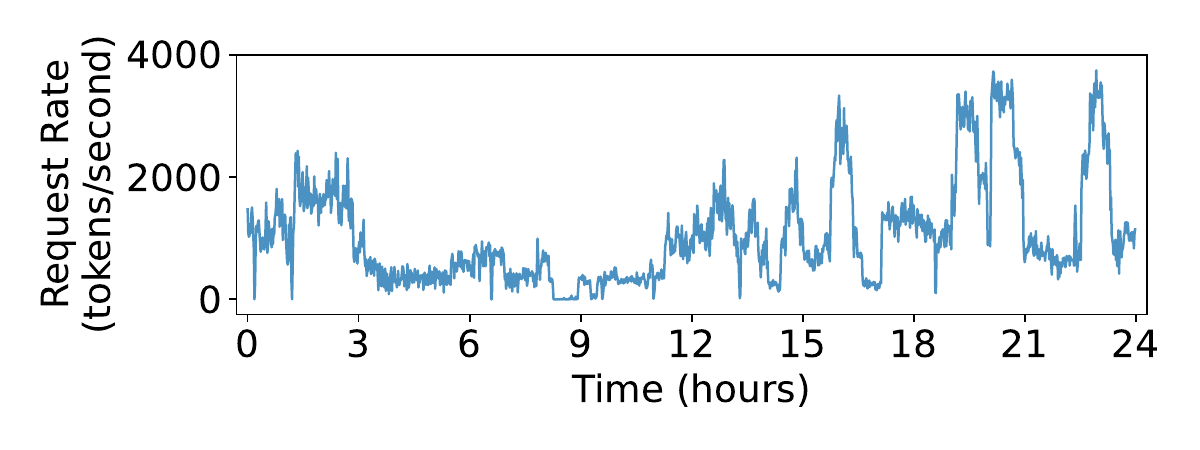}
        \vspace{-1.em}
        \caption{Load variation over 24 hours.\label{fig:motiv:burstgpt-a}}
    \end{subfigure}\\
    \begin{subfigure}[b]{0.9\linewidth}
        \centering
        \includegraphics[width=0.99\linewidth]{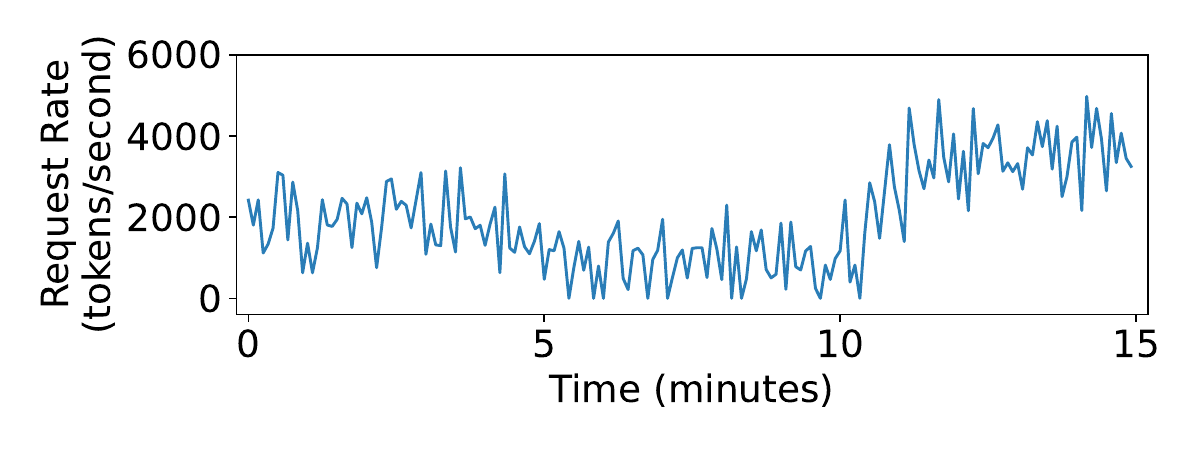}
        \vspace{-1.em}
        \caption{Load variation over 15 minutes.\label{fig:motiv:burstgpt-b}}
    \end{subfigure}
    \caption{User traffic to ChatGPT within a campus exposes high load variability at various time scales.\label{fig:motiv:burstgpt}}
\end{figure}

\mysection{Motivation}
\label{sec:motivation}

In this section, we first demonstrate how real-world LLM load can vary over time and why existing serving systems fall short of achieving high GPU utilization. We then use an experiment to quantitatively demonstrate the limitations of simply co-locating online and offline serving with preemptive scheduling.

\mysubsection{GPU Underutilization in LLM Serving}
\label{sec:gpu_underutilize}
\MyPara{Online Load Burstiness.}
Real-world LLM workloads often expose diurnal patterns and high load variability, as shown by reports from LLM service providers~\cite{mooncake-kv,patel2024splitwise} and a recent study~\cite{burstgpt} which collects ChatGPT user traffic for two months. Figure~\ref{fig:motiv:burstgpt}(a) presents the sampled request rate within a day. Despite the average load being moderate, there is a clear contrast between peak hours in the afternoon and non-peak hours in the morning.
Beyond diurnal patterns, this load variance also extends to shorter timescales, and the request rate can ramp up multiple times within minutes.
Figure~\ref{fig:motiv:burstgpt}(b) zooms in on a 10-minute window to analyze one such burst.
As shown, the load still fluctuates drastically at the minute timescale, and the request rate increases by 3$\times$ at time $t=10$min. 
Traces from other LLM service providers, such as Azure, also expose similar high load variability~\cite{patel2024splitwise}.

\begin{figure}[t]
    \centering
    \includegraphics[width=0.85\linewidth]{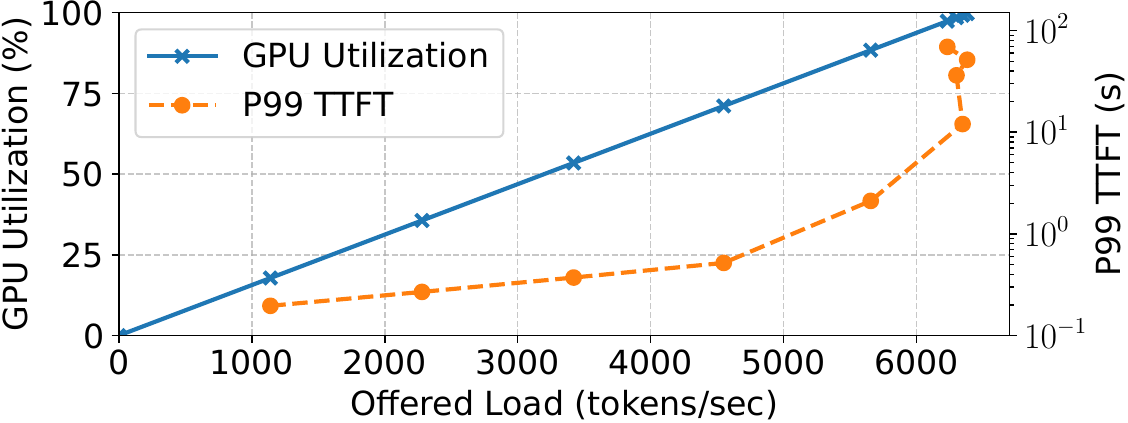}
    \caption{GPU utilization and the P99 TTFT across request rates measured with Llama-3.1 8B on an H100 GPU. Utilization is defined as the offered load relative to the GPU’s maximum achievable load.}
    \label{fig:motiv:gpu-util}
\end{figure}

\MyPara{Static Reservation Underutilizes GPUs.}
LLM service providers often use long-term reserved or on-premise GPU instances, which both mitigate GPU unavailability and offer greater cost efficiency than on-demand instances~\cite{xia2025skylb}.
Due to such high load variability, online serving must reserve sufficient resources for the peak demand to avoid violating its stringent latency SLOs. However, since the peak inference load can be multiple times higher than the average load, overprovisioning can lead to significant resource waste.
Figure~\ref{fig:motiv:gpu-util} illustrates the P99 TTFT and GPU utilization under different request rates for a Llama-3.1 8B model running on an H100 GPU.
As shown, the model achieves a good balance between high GPU utilization and acceptable tail TTFT at a load of 5000 tokens/s, which is roughly the peak load in Figure~\ref{fig:motiv:burstgpt-b}. However, under a moderate load of 2000 tokens/s (average load in Figure~\ref{fig:motiv:burstgpt-b}), GPU utilization drops sharply, and over 70\% of resources are left idle.

\mysubsection{Challenges in LLM Co-Serving}
Co-serving online and offline workloads on the same LLM serving instance is a promising way to improve GPU utilization. 
However, co-serving faces a fundamental challenge in balancing the low-latency requirements of online inference with the high-throughput demands of offline inference.

To quantify this trade-off, we extended a state-of-the-art serving engine vLLM~\cite{vllm-piecewise-cuda-graph} with a priority-based scheduler and preemption support (details in \S\ref{sec:eval:setup}). At a high level, the scheduler builds on chunked-prefill scheduling~\cite{sarathi-serve@arxiv24} by adding additional offline tokens when online requests are unable to fill up the chunk. 
Upon receiving online requests, our preemptive scheduler, denoted as \codeIn{Preemptive}, pauses running offline requests. It keeps their KV cache in GPU memory as long as space permits, deferring eviction and avoiding recomputation until memory pressure necessitates it. 
Meanwhile, to estimate the upper bound of achievable offline throughput, we also introduce a non-preemptive variant, \codeIn{Non-Preemptive}, which still schedules offline requests at a low priority but allows scheduled requests to run to completion.

\begin{figure}[t]
    \centering
    \small
    \includegraphics[width=0.99\linewidth]{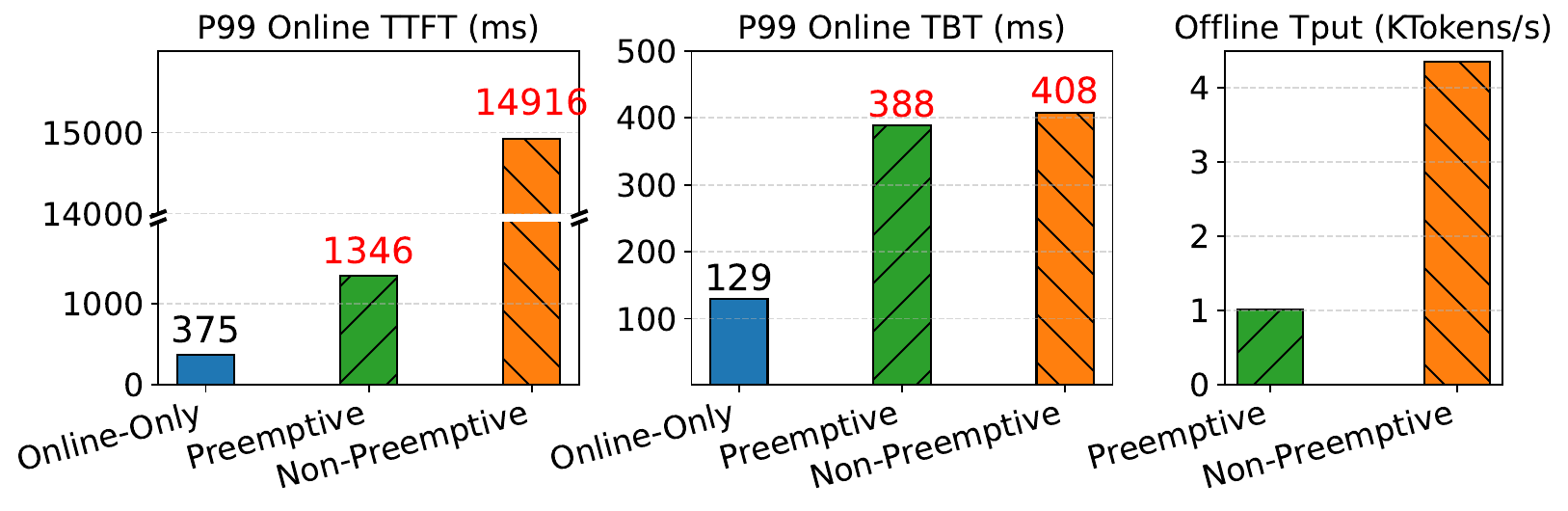}
    \vspace{-1em}
    \caption{P99 online latencies and offline throughput when co-serving online and offline requests with preemptive or non-premptive scheduling with Llama-3.1 8B on one H100 GPU.
    \label{fig:motiv:naive-colocation} }
\end{figure}

We compared these two policies against \codeIn{Online-Only}, which serves only online inference and represents optimal online latency, on an H100 GPU serving Llama-3.1 8B. We used a synthetic online workload (Gamma process, CV=0.5, 2 req/s) alongside sufficient offline requests. Each request has 4096 input tokens and 256 output tokens (see \S\ref{sec:eval:load-varation} for other request lengths). 
Figure~\ref{fig:motiv:naive-colocation} reports the P99 online TTFT, TBT, and offline throughput.
As shown, simple preemptive scheduling can achieve neither low online latency nor high offline throughput. 
Using the preemptive scheduler increases P99 online TTFT by 3.59$\times$ and TBT by 3.01$\times$. 
Meanwhile, the preemptive scheduler reduces offline throughput by 4.27$\times$ compared to non-preemptive scheduling. Our analysis revealed three root causes:

\squishlist
\item \textbf{Coarse-grained scheduling leads to contention}: With chunked-prefill, the scheduler operates at the chunk granularity. To maximize utilization, it greedily fills each chunk with offline tokens but ignores the potential compute and memory contention they introduce. This directly slows down the decoding of concurrent online tokens and inflates online TBT.

\item \textbf{Iteration-level preemption causes queueing delays}: Preemption occurs only after a full iteration finishes, meaning incoming online requests must wait for the entire in-flight chunk to complete. This coarse granularity creates head-of-line blocking that can last hundreds of milliseconds, severely degrading TTFT.

\item \textbf{Request-level KV cache management incurs high overhead}: When preemption occurs, state is managed at the entire sequence's KV cache granularity. This forces a costly dilemma: either recompute the entire KV cache later or stall the GPU to swap it to host memory. Our profiling shows that under memory pressure, up to 69\% of GPU time is wasted on recomputation or swapping~\cite{vllm@sosp23}.   
\squishend

\MyPara{Takeaway.} 
These challenges necessitate a system featuring fine-grained resource management: (1) Adaptive scheduling that bounds latency impact of co-served offline work within online SLOs; (2) sub-iteration preemption to mitigate head-of-line blocking; and (3) sub-request KV cache management for efficient context switching.

\mysection{Design}
\label{sec:design}

\begin{figure}[t]
    \centering
    \small
    \includegraphics[width=0.95\linewidth]{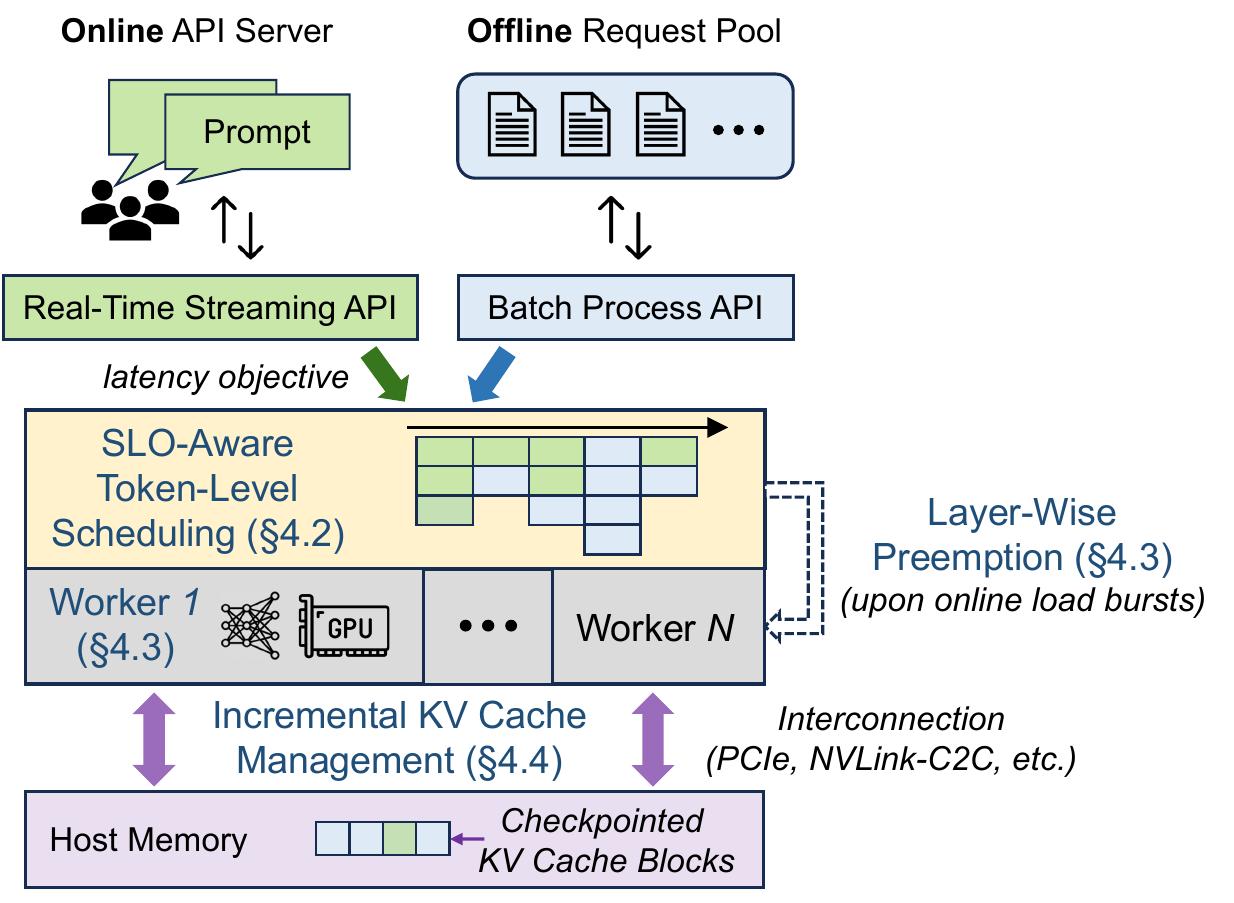}
    \caption{Design overview of \tool. 
    }
    \label{fig:overall-design}
\end{figure}

\mysubsection{\tool Overview}
\label{sec:design:overview}

As shown in Figure~\ref{fig:overall-design}, \tool is an LLM serving system that co-serves online and offline requests to maximize utilization while preserving online SLOs and offline progress. The key to its efficiency is breaking the coarse-grained limitations of existing systems and utilizing resources at a much finer granularity across three core components of the serving process: scheduling, model execution, and KV cache management.
\squishlist
\item \textbf{SLO-aware token-level scheduling} (\S\ref{sec:design:scheduler}). 
\tool's scheduler admits requests at the token granularity. Instead of using a static, fixed-size chunk, the scheduler employs a context-aware performance model to calculate a precise token budget for every single iteration. This allows it to safely fill offline tokens for maximized utilization subject to SLO and adapt in real-time to workload dynamics. 
\item \textbf{Layer-wise preemption} (\S\ref{sec:design:preemption}). To provide immediate responsiveness to online traffic, \tool's execution engine supports sub-iteration preemption. By instrumenting the model, it creates preemption points between model layers, allowing the scheduler to interrupt offline work in single-digit milliseconds. This breaks the atomicity of the iteration, enabling the system to react to SLO threats at the layer level rather than waiting for a full, potentially long-running, forward pass to complete.
\item \textbf{Incremental, token-level state management} (\S\ref{sec:design:inc-checkpointing}). To make fine-grained preemption practical, \tool manages the KV cache with incremental, token-level checkpointing. Leveraging the insight that the cache is append-only, it asynchronously saves only the state for the single, newly-generated token after each iteration. This decouples the cost of preemption from a request's history, making the overhead constant and negligible, and resolving the costly swap-vs-recompute dilemma of monolithic, sequence-level state management.
\squishend

\mysubsection{SLO-Aware Token-Level Scheduling}
\label{sec:design:scheduler}

In this section, we first study why existing schedulers with static budgets are ``unsafe''
for co-serving, \ie, they can frequently cause online SLO violations. We then introduce our context-aware performance model and present a dynamic, token-level scheduling algorithm that uses this model to safely maximize GPU utilization while adhering to strict online SLOs.

\MyPara{Limitations of static chunk-level scheduling.}
Chunked prefill~\cite{sarathi-serve@arxiv24} stands for the state-of-the-art scheduler for online serving and has been enabled by default in modern serving engines~\cite{sglang@arxiv23,vllm@sosp23}. It splits a long prefill into smaller chunks and schedules them alongside decode phases across iterations, which reduces head-of-line blocking from long prefill and stabilizes TBTs. 
The algorithm assumes that a chunk of $k$ tokens takes roughly constant time to execute and uses a fixed chunk size as the budget for the number of tokens to compute.

While this assumption may suffice in online serving when GPUs are rarely saturated, it is unsafe for co-serving when GPUs run at a high utilization and are sensitive to latency. This is because it ignores the performance impact of the context and KV caches being loaded for all active requests and can lead to inaccurate latency estimations. 
For instance, prefilling a 2048-token chunk for Llama-3.1 8B on an H100 takes 51ms for a new request but increases to 124ms when the request already has a 40K token context.
Given a 100ms TBT SLO, a static chunking policy treats both scenarios identically and makes the same scheduling decision, leading to an SLO violation in the latter case.
This demonstrates that no static chunk size can simultaneously achieve both high efficiency and strict latency guarantees across varying request contexts.

\MyPara{Context-aware performance model.}
To accurately predict the latency of any potential batch, we first build a context-aware performance model to analyze the performance of autoregressive transformer-based LLMs.
We model the latency as a function of two key variables: the number of compute tokens ($P$) being actively processed, and the number of pre-existing context tokens ($C$) held in the KV cache. 

The latency for a given iteration is approximated by the following polynomial:

\begin{align*}
\textbf{Latency}(P, C) &= \underbrace{k_1 P}_{\text{Linear Compute}} +\quad \underbrace{k_2 P(P + C)}_{\text{Attention}} \\
&+ \underbrace{k_3 P}_{\text{Communication}} + \underbrace{k_4 (P + C)}_{\text{Memory Access}} + \underbrace{k_5}_{\text{Constant}}
\end{align*}

Each term in the model captures a distinct aspect of the transformer architecture's performance:
\begin{enumerate}
    \item \textbf{Linear compute} ($k_1 P$): This term covers per-token operations like the QKV and MLP projections, whose cost scales linearly with the number of compute tokens.
    \item \textbf{Attention} ($k_2 P(P+C)$): This is the most critical term for context-awareness. It models the quadratic complexity of attention, which depends on both the new tokens being processed ($P$) and their interaction with all existing context tokens ($C$). This term often dominates latency in requests with long contexts.
    \item \textbf{Communication} ($k_3 P$): This captures the overhead of collective communication operations (e.g., \codeIn{all reduce}) in tensor parallel configurations, which scales with $P$ being communicated.
    \item \textbf{Memory Access} ($k_4 (P+C)$): This models the cost of memory-bound operations, primarily loading the KV cache from HBM for the attention computation.
    \item \textbf{Constant} ($k_5$): This accounts for fixed overheads in each iteration, such as loading model weights and other small, constant-time kernels.
\end{enumerate}

\tool learns the conefficients ($k_1$ to $k_5$) via a one-time, offline profiler. For a given model and hardware setup, the profiler empirically measures iteration latency across a grid of $P$ and $C$ values and uses polynomial regression to fit the model.
In practice, the profiling process is fast and accurate. It finishes in 20 minutes even for 70B models with a relative prediction error below 4\%.

The profiler offers a \codeIn{can\_schedule(batch,req,TBT)} interface that informs the scheduler if a new request can be added into the batch while adhering to TBT SLO. It returns the number of compute tokens that can be scheduled from \codeIn{req} or 0, which means \codeIn{req} cannot be put into the batch with the current TBT objective.

\MyPara{Latency-aware token-level scheduling.} 
Armed with the predictive model, the \tool scheduler executes a dynamic, token-level algorithm at every iteration to compose the optimal batch\footnote{This paper uses ``batch'' and ``chunk'' interchangeably; in chunked prefill, a batch refers to a chunk of compute tokens from scheduled requests.}.
At a high level, the scheduler prioritizes online requests
to avoid introducing any queuing delay. 
After accommodating all pending online requests, it iterates through offline requests and leverages the profiler to decide how many offline requests and tokens can be added into the batch given the online TBT SLO.

In addition to latency, the scheduler must also manage the finite capacity of the GPU's KV cache. A unique challenge in co-serving is that idle offline requests, which are paused to meet online SLOs, still consume memory and block incoming online requests. To resolve this, \tool releases their KV caches on demand to incorporate online requests. We will discuss how \tool manages and restores KV caches efficiently in \S\ref{sec:design:inc-checkpointing}.

Finally, during periods of no online traffic, the scheduler enters an offline-optimized mode. It uses the performance model to determine the optimal batch size that maximizes throughput until the next online request arrives.

\mysubsection{Layer-Wise Preemption} 
\label{sec:design:preemption}

Real-world workloads often see sudden online load changes (\S\ref{sec:gpu_underutilize}). However, a single iteration of a large offline batch can occupy the GPU for up to seconds, blocking the execution of incoming online requests and potentially causing online TTFT violations.
To address this, online requests must be able to preempt offline requests at a sub-iteration level.

However, this is challenging because, for efficiency, inference engines dispatch a batch to dedicated GPU workers to execute an iteration uninterruptibly as a single executable or a CUDA Graph and return control only when the iteration finishes. Intra-iteration interruption can leave inconsistent states and potentially crash the engine.

\tool breaks this iteration-level barrier by instrumenting the model at the Transformer layer granularity, which brings two key advantages. First, modern LLMs typically comprise dozens of layers where each layer runs for a few milliseconds, ensuring high responsiveness. Second, layers represent natural computation boundaries with clear input and output dependencies, which allows offline requests to pause and exit early with simple cleanup.

\begin{algorithm}[t]
\caption{Preventing online TTFT violations.\label{alg:preempt-on-online}}
\small
\DontPrintSemicolon
\SetKwComment{Comment}{$\triangleright$\ }{}

\SetKwFunction{OnRecvOnlineRequest}{\textsc{OnRecvOnlineRequest}}
\SetKwFunction{PreemptRunning}{\textsc{PreemptRunning}}
\SetKwProg{Fn}{Function}{:}{}
Let $Q_{on}$ denote incoming online requests, and $B$ denote the current batch.\;
\KwIn{(1) Online TTFT objective $t_{TTFT}$.}
\Comment{Run in the monitor thread.}
\Fn{\OnRecvOnlineRequest{$t_{TTFT}$}}{
    \If{no offline requests in the current batch}{
        \Return
    }

    $t_{on} \gets$ Profiler.estimate\_exec\_time($Q_{on}$)\;\label{alg:line:estimate-online}

    $t_{\textit{remaining}} \gets$ Profiler.estimate\_exec\_time($B$) - $t_{passed}$ \; \label{alg:line:estimate-offline}
    \If{$t_{\textit{remaining}} + t_{on} > t_{\textit{TTFT}}$}{ \label{alg:line:preempt-running-stt}
        Worker.signal\_layerwise\_preemption()\;\label{alg:line:preempt-running-func}
    } \label{alg:line:preempt-running-end}
}   
\end{algorithm}

\tool codesigns the scheduler and workers to realize layer-wise preemption.
The scheduler is responsible for detecting potential latency violations. It spawns a dedicated monitor thread, which is triggered upon receiving a new online request. Algorithm~\ref{alg:preempt-on-online} outlines its logic. It uses the performance model and profiler (\S\ref{sec:design:scheduler}) to estimate if waiting for the current batch to finish would cause the new request to miss its TTFT SLO. If a violation is predicted, the scheduler immediately signals for preemption by writing to a designated flag in shared host memory.

\tool workers instrument the model with lightweight synchronization points (safepoints) between layers, inspired by language runtime designs~\cite{hotspotvm-safepoint}. At each safepoint, the worker checks the preemption flag with a single device-to-host memory load. If the flag is set, the worker will terminate computation for all offline requests in the batch, clean up their hidden states and attention metadata, and then resume execution for the remainder of the iteration with only the online requests.

\MyPara{Tensor parallelism.} One additional challenge in implementing the safepoint with tensor parallelism (TP) is that collective communication (\eg, all-reduce) in TP requires strict synchronization across workers. Arbitrary preemption of a subset of workers can cause communication deadlocks~\cite{pan2023occl}. To address this, \tool will synchronize all workers with \codeIn{ncclBroadcast} at the beginning of the safepoint before checking for preemption signals.

\MyPara{CUDA graph compatibility.} Our safepoint mechanism is fully compatible with CUDA Graphs. First, modern engines such as vLLM capture each model layer as a separate graph, where we can insert the safepoint between graph launches. For systems that capture the model as a single graph, safepoint can be implemented as a conditional CUDA graph node~\cite{nvidia-conditional-cuda-graph} within the graph.

\MyPara{Overhead analysis.} We have carefully implemented safepoint to minimize its cost. Each safepoint check consists of a single read from pinned host memory (7\us) and a single broadcast/barrier operation across GPUs (13\us on NVLink). This total cost of 21\us is negligible compared to the several milliseconds required to execute a transformer layer. More details are presented in \S\ref{sec:eval:drilldown:preemption-cost}.

\begin{figure}[t]
    \centering
    \small
    \includegraphics[width=0.99\linewidth]{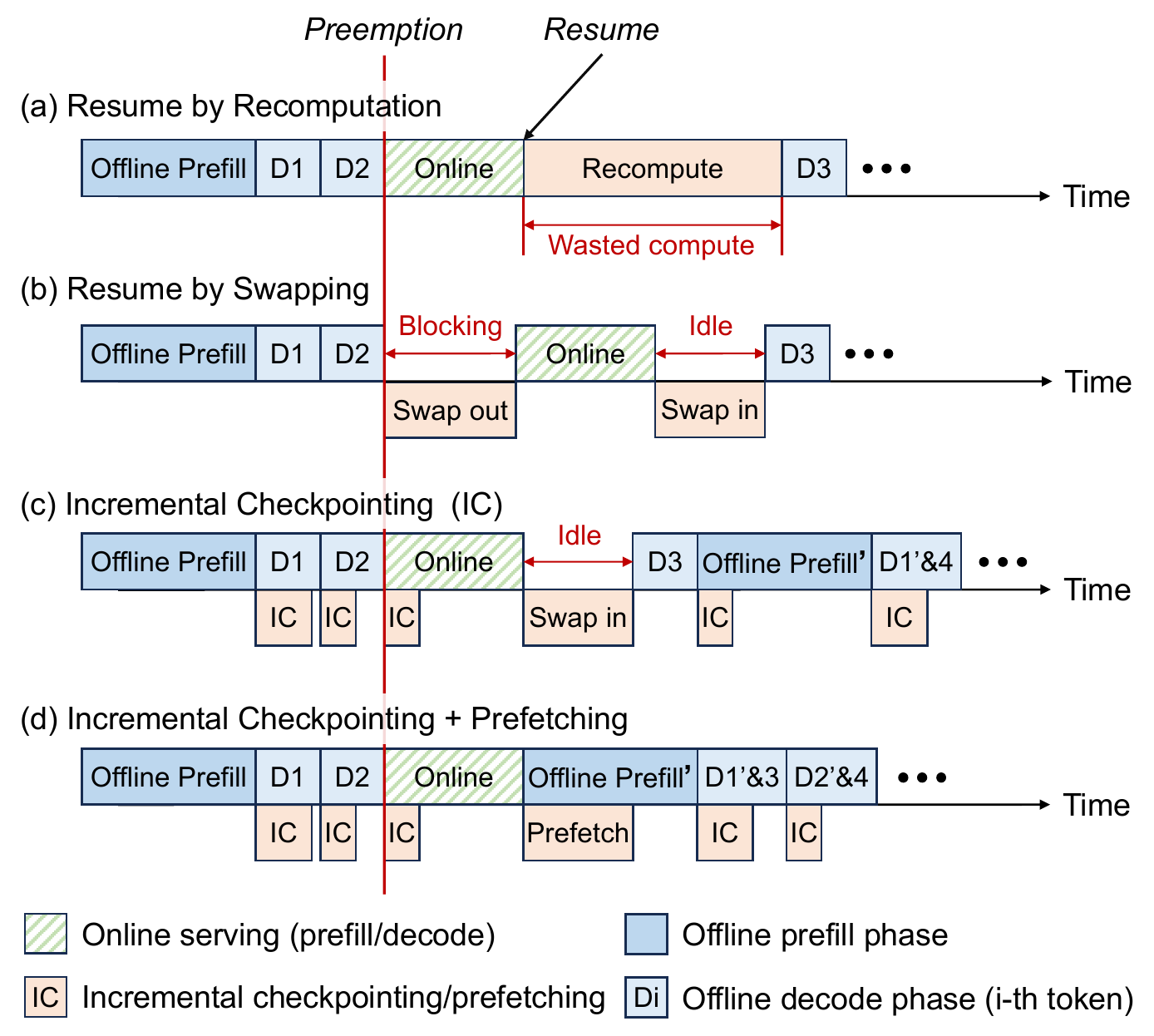}
    \caption{Different preemption and resumption strategies. (a) Resuming by recomputation achieves low preemption delay at the cost of additional computation. (b) Resuming by swapping avoids recomputation but blocks the schedule of incoming online requests. (c) Incremental checkpointing (IC) minimizes both preemption delay and resumption cost. (d) Incremental checkpointing + prefetching overlaps swap-in with computation of the next batch.
    \label{fig:inc-checkpointing}}
\end{figure}

\mysubsection{Incremental KV Cache Management}
\label{sec:design:inc-checkpointing}

Effective preemption requires not only interrupting computation but also rapidly reclaiming GPU memory. While existing systems, including SGLang~\cite{sglang@arxiv23} and vLLM~\cite{vllm@sosp23}, allocate KV cache memory in fine-grained chunks (tokens~\cite{sglang@arxiv23} or pages~\cite{vllm@sosp23}), they \textit{evict them at request granularity} during preemption: the KV history of a preempted request is treated as a single, indivisible unit. Consequently, the entire KV of that request is evicted at once.

This coarse-grained ``fate-sharing'' forces a costly tradeoff between reclamation latency and restoration cost. Figure~\ref{fig:inc-checkpointing} demonstrates this with two strawman eviction approaches: (a) recomputation and (b) swap.
Recomputation immediately discards the KV cache but wastes GPU cycles to recompute the offline job from scratch later.
Swapping preserves KV cache states by writing them to host memory, but the blocking I/O delays the scheduling of incoming online requests proportionally to the number of tokens of the request, causing SLO violations.

\MyPara{The fine-grained solution: decoupling lifecycles.}
Such inefficiency stems from the tight coupling of a request's lifecycle and its KV caches.
\tool tackles this by managing the KV cache state of each token individually. This fine-grained design allows tokens and their KV caches of a preempted request to be flexibly distributed across GPU, host memory, or marked for recomputation.

\MyPara{Incremental checkpointing and prefetching.}
Enabled by per-token state management, \tool can incrementally checkpoint and load KV cache of each token independently with their parent request. 
Leveraging the immutability of KV cache (\ie, append-only), \tool proactively checkpoints KV caches of only the newly generated tokens to host memory after each generation step (Figure~\ref{fig:inc-checkpointing}(c)). This amortizes the I/O traffic across multiple iterations and avoids sudden bursts of device-to-host data transfers, allowing I/O to fully overlap with computation.

Unlike traditional swapping, incremental checkpointing keeps KV caches in GPU memory rather than discarding them directly, so ongoing offline requests can continue inference until preemption occurs, at which point their KV caches will be discarded on demand at the token level.

Symmetrically, when GPU memory is available for resuming a preempted request, \tool prefetches KV caches for evicted tokens incrementally in the background. As shown in Figure~\ref{fig:inc-checkpointing}(d), prefetching I/O is overlapped with the computation and completely off the critical path.

\MyPara{Cost and feasibility analysis.}
The time overhead of incremental KV checkpointing is negligible. On an H100 GPU with 64GB/s device-to-host PCIe interconnect, data transfer throughput for Qwen-2.5 14B (192KB KV cache per token) achieves $64 \times 1024^2/ 192=349$\textbf{K}tokens/s, which is much higher than LLM generation throughput. Our measurement shows that even transferring an entire chunk of 2048 tokens takes only 10ms and can be effectively hidden by computation overlapping (detailed evaluation in \S\ref{sec:eval:drilldown}).

The host memory footprint is also strictly bounded by the total physical GPU memory. In a typical setup where host memory is overprovisioned (\eg, DGX server with 8 H100s, 640 GB of GPU memory, and 2 TB of host memory), this is not a practical constraint.
Even under extreme cases with limited host memory, \tool can gracefully degrade and adapt to the available host memory size. When host memory is exhausted,
\tool can evict individual tokens and fall back to recomputation for only those specific tokens, which is still more efficient than a coarse-grained, all-or-nothing eviction policy.

\mysection{Implementation}

We have implemented \tool atop vLLM 0.8.4~\cite{vllm@sosp23}.
by adding or modifying 9169 lines of Python and C++ code.

To support incremental KV cache checkpointing, \tool manages KV cache at a granularity of 16 tokens to align with vLLM page size, and enhances vLLM's KV cache manager to track the mapping between each GPU KV page and its corresponding CPU KV page for checkpointing. This mapping is recorded in an extended field of the virtual page table.
To support efficient KV cache transfer for tokens scattered in GPU memory, we implement an efficient kernel that gathers their KV cache into a contiguous buffer and transfers it once.
To ensure overlap between computation and asynchronous swap I/O, \tool creates a separate CUDA stream for KV cache checkpointing operations. Original model executable or CUDA Graphs are still launched in the default stream as is.
Since the checkpointed or prefetched KV cache blocks have no data dependencies with the ongoing model computation, no synchronization is required between the two CUDA streams.

\tool includes a standalone offline profiler that can be run independently with model configuration to generate a performance profile, which is then loaded by the scheduler at initialization.
\tool also comes with a built-in load generator that can generate precisely timed requests following the gamma distribution. The load generator can be configured with various parameters including the request rate, burstiness (\ie, skewness of the gamma distribution), and request lengths to cover the characteristics of real-world workloads, enabling users to better understand and optimize \tool's performance.

\mysection{Evaluation}

Our evaluation aims to answer the following questions:
\begin{enumerate}[nosep, leftmargin=*]
\item Can \tool efficiently harvest GPU resources while maintaining low online latency? (\S\ref{sec:eval:synthetic-overall})
\item Can \tool achieve both low online latency and high offline throughput under real-world workloads? (\S\ref{sec:eval:real-overall})
\item Can \tool maintain efficiency for workloads with different degrees of burstiness and request lengths? (\S\ref{sec:eval:load-varation})
\item How does each component contribute to \tool's efficiency?
(\S\ref{sec:eval:drilldown})
\end{enumerate}

\mysubsection{Setup}
\label{sec:eval:setup}
\MyPara{Environment and models.} 
We conducted experiments on a DGX server with two 48-core CPUs, 2TB of memory, and eight NVIDIA H100 GPUs connected with 900GB/s fully-meshed NVLink.
The server runs Ubuntu 22.04 and CUDA 12.8. To reduce latency jitter, we disabled dynamic voltage and frequency scaling (DVFS) of GPUs~\cite{gpu-dvfs@eenergy19, nvidia-smi-doc} and Python garbage collector~\cite{python-gc-doc}. We have verified that these settings do not negatively impact LLM inference performance.

We evaluated \tool on three models of varying sizes, including Llama-3.1 8B~\cite{grattafiori2024llama3herdmodels}, Qwen-2.5 14B~\cite{qwen2.5techreport@arxiv}, and Llama-3.1 70B~\cite{grattafiori2024llama3herdmodels}. 
We served each model with FP16 precision and a minimum number of GPUs, which is 1 H100 for Llama-3.1 8B and Qwen-2.5 14B, and 4 H100s for Llama-3.1 70B with tensor parallelism. Among them,
Llama-3.1 8B represents a resource-rich setup with enough GPU memory for KV caches (16GB model weights, 60GB KV cache),
Qwen-2.5 14B reflects a more constrained setup with 28GB of model weights and 48GB allocated for KV cache, and
Llama-3.1 70B demonstrates tensor parallelism across multiple GPUs.

\MyPara{Baselines.}
We compared \tool with four baselines: 
\begin{enumerate}[
leftmargin=*, label=(\arabic*)]
\item \textit{Online-Only (Latency Optimal)} refers to the original vLLM configuration, which serves only online inference. It provides optimal online serving latency but zero offline throughput.
\item \textit{Non-Preemptive (Throughput Optimal)} is a variant policy designed to maximize offline throughput. It prioritizes online requests over offline ones, but schedules offline requests at a lower priority when online inference does not fully utilize the GPU, but does not preempt running offline requests when new online requests arrive. While this approach degrades online serving latency, it provides an upper-bound estimate of offline throughput by allowing offline requests to complete uninterrupted.
\item \textit{Sarathi-Preemptive (Sarathi-P)} incorporates Sarathi’s~\cite{sarathi-serve@arxiv24} chunked-prefill scheduling. To create a strong co-serving baseline, we extend its scheduler to prioritize online requests and then opportunistically fill offline tokens with the remaining chunk space. The scheduler pauses offline requests when the token budget is met, and evicts their KV caches only under memory pressure.
\item \textit{DistServe-Preemptive (DistServe-P)} represents the disaggregated architecture~\cite{distserve@arxiv24}, where prefill and decode operations are handled by separate sets of GPUs. We extend its scheduler similar to (3) on both the prefill engine and decode engine to prioritize online requests and preempt offline requests under memory pressure.
We carefully tuned the ratio of prefill-to-decode GPUs (e.g., 2:1, 1:1, 1:2) and report the best-performing configuration. 

\end{enumerate}
We implemented or ported all baselines to vLLM 0.8.4~\cite{vllm@sosp23} with chunked-prefill enabled by default
for fair comparisons.
For all systems, configurations were carefully tuned for their optimal performance.

\begin{table}[t]
    \centering
    \begin{adjustbox}{max width=0.99\linewidth}
    \aboverulesep=0ex
    \belowrulesep=0ex
    \begin{tabular}{c|c|rr|rr}
    \toprule\rule{0pt}{1.1EM}%
    \multirow{2}{*}{\textbf{Task}} & \multirow{2}{*}{\textbf{Dataset}} & \multicolumn{2}{c|}{\textbf{Input Tokens}} & \multicolumn{2}{c}{\textbf{Output Tokens}} \\
    &  & Mean & P99 & Mean & P99 \\
    \midrule\rule{0pt}{1.1EM}%
    Online chatting & BurstGPT~\cite{burstgpt}  & 2747 & 3446 & 267 & 481 \\
    \midrule\rule{0pt}{1.1EM}%
    Benchmarking & DuReader~\cite{he2018dureaderchinesemachinereading} & 21776 & 29306 & 166 & 754 \\
    Question answer & MultiNews~\cite{fabbri-etal-2019-multinews} & 6916 & 12084 & 394 & 788 \\
    Summarization & VCSUM~\cite{wu2023vcsumversatilechinesemeeting} & 18803 & 32768 & 337 & 510 \\
    \bottomrule
    \end{tabular}
    \end{adjustbox}
    \caption{Real-world online and offline workloads used for evaluation.
    \label{tab:workloads-overview}}
\end{table}

\MyPara{Real-world workloads.}
For the online workload, we evaluated BurstGPT~\cite{burstgpt}, a representative trace of user requests to GPT-4~\cite{openai2023gpt4} collected from a university campus. 

For offline workloads, we evaluated three tasks: DuReader \cite{he2018dureaderchinesemachinereading} for model benchmarking, MultiNews~\cite{fabbri-etal-2019-multinews} for multi-document summarization, and VCSUM~\cite{wu2023vcsumversatilechinesemeeting} for meeting summarization.
Table~\ref{tab:workloads-overview} summarizes all four evaluated workloads, including their input and output length distributions.

\MyPara{Synthetic workloads.} 
To evaluate \tool{}'s ability to balance efficiency with strict SLOs under diverse conditions, we generate synthetic request arrivals using a Gamma process, following methodologies established in prior trace studies~\cite{burstgpt,fairnessinllm@osdi24,alpaserve@osdi23}.
We vary key parameters, including request rate, SLO tightness, request lengths, and traffic burstiness for overall performance testing (\S\ref{sec:eval:synthetic-overall}) and sensitivity analysis (\S\ref{sec:eval:load-varation}).

\MyPara{Metrics.} Since online and offline serving have different performance objectives, we evaluate them with distinct metrics. For online serving, following prior work~\cite{sarathi-serve@arxiv24}, we measure each request’s P99 TTFT and TBT. For offline serving, we evaluate throughput by measuring the number of tokens processed per second, including both prefill and decode tokens.

\begin{figure*}[t]
    \centering
    \includegraphics[width=0.93\linewidth]{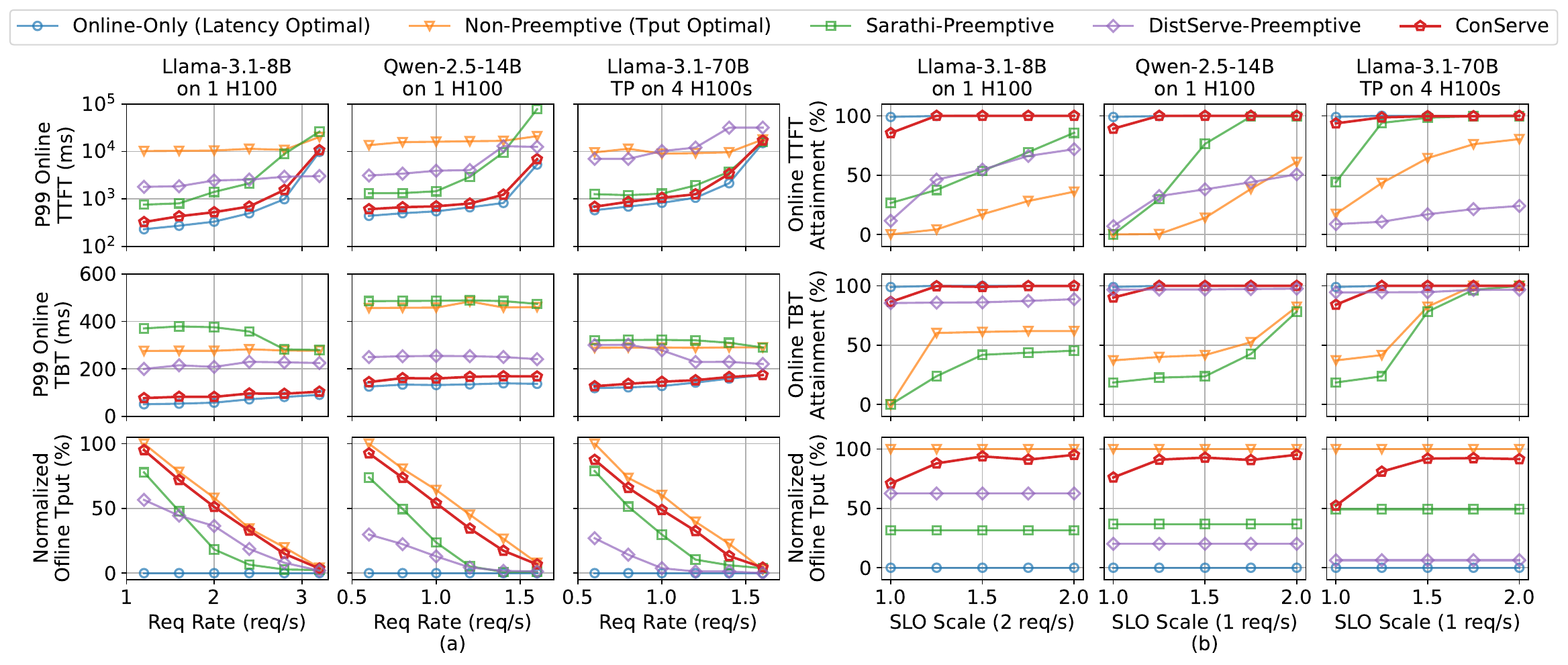}
    \small
    \caption{Overall serving performance for all three models on synthetic workloads with varying (a) request rates and (b) SLO scales.
    \tool achieves consistently low TTFT and TBT that are comparable with optimal online latency (measured with \codeIn{Online-Only}) and high overall serving throughput that are close to optimal offline throughput (measured with \codeIn{Non-Preemptive}).
    }
    \label{fig:eval:synthetic-overall-perf}
\end{figure*}

\mysubsection{GPU Efficiency and Latency}
\label{sec:eval:synthetic-overall}

This subsection investigates whether \tool can efficiently harvest idle GPU resources\textemdash whenever the online request rate is low or the latency SLO is sufficiently relaxed\textemdash for offline throughput while still preserving online SLOs.

\MyPara{Setup.}
To enable a comprehensive evaluation, we generate synthetic online loads with varying request rates and latency SLO scale. Following workload parameters from BurstGPT~\cite{burstgpt}, we set the burstiness (measured as the coefficient of variation, CV) to 0.5, with each request containing 4096 input tokens and 256 output tokens.
A sensitivity analysis on request lengths and burstiness is provided in \S\ref{sec:eval:load-varation}.

We configured \tool's target SLOs based on the P99 latencies achieved by \codeIn{Online-Only}.

\MyPara{Online load vs. performance trade-offs.}
Figure~\ref{fig:eval:synthetic-overall-perf}(a) shows the P99 online TTFT, TBT, and offline throughput for all systems as the online request rate increases across all models. The results demonstrate that \tool is the only system that achieves both low online latency and high offline throughput.

Regarding online latency, \tool maintains P99 TTFT and TBT close to the optimal online latency baseline (\codeIn{Online-Only}) across all tested request rates, staying within 25\% and 19\% on average, respectively.
In contrast, the other baselines suffer from significant latency degradation due to their coarse-grained designs.
\codeIn{Non-Preemptive} suffers from high TTFT and TBT due to head-of-line blocking caused by scheduled offline requests.
Two preemptive baselines improve TTFT but remain slow in TBT because they cannot bound the compute and memory interference introduced by offline tokens. Notably, \codeIn{Sarathi-P} can sometimes yield worse TBT than \codeIn{Non-Preemptive} because preemption forces costly recomputation of the entire prefill phase.

\codeIn{DistServe-P} shows poor latencies because its disaggregated architecture creates load imbalance between prefill and decode instances which amplifies the contention from offline requests. Decode instances quickly exhaust their GPU memory and become the bottleneck due to excessive memory-heavy decoding requests. This triggers frequent swapping that not only stalls decoding but also applies back-pressure to the prefill instances, ultimately degrading both TBT and TTFT.

In terms of offline throughput, \tool consistently outperforms all other preemptive baselines and achieves an average of 82.3\% of the optimal throughput.
At low request rates, where GPU memory is underutilized, both \codeIn{Sarathi-P} and \tool maintain over 70\% of optimal offline throughput, as GPU memory is sufficient to keep offline requests' KV caches.
However, as load increases, \tool’s advantage becomes more pronounced. By eliminating the recomputation overhead, \tool achieves an average of 3.0$\times$ (up to 7.48$\times$) higher offline throughput compared to \codeIn{Sarathi-P}.
\codeIn{DistServe-P} yields the lowest throughput in most cases due to frequent preemptions on decode instances and back-pressure on prefill instances.

\MyPara{Online SLO scale vs. performance trade-offs.}
Figure~\ref{fig:eval:synthetic-overall-perf}(b) illustrates how system performance changes as we relax the online SLO scale $s$, where $\text{SLO}_{\text{TTFT/TBT}}(s)=s\cdot P99_{\text{TTFT/TBT}}$, dictating the allowed latency increase over \codeIn{Online-Only}.

Uniquely, \tool can trade-off offline throughput for stricter online SLO adherence, enabling >99\% TTFT and TBT attainment in all settings when $s\!>\!1$. 
Even in the most challenging setting with no slack ($s=1.0$), where minor system jitter can cause violations, \tool still maintains at least 86\% attainment, significantly exceeding all baselines.
This SLO awareness also enables \tool to harness latency slack for higher offline throughput until the $s=1.5$, where \tool achieves an average 93\% optimal throughput. Further increasing $s$ brings a diminishing 2\% throughput improvement.

In contrast, all other baselines are oblivious to the configured SLO scale, and their offline throughput remains constant.
\codeIn{DistServe-P} presents an interesting case. It can achieve high TBT attainment thanks to the disaggregation. But the attainment is still lower than \tool in most cases because preemptions still cause TBT jitter. Its TTFT attainment remains poor due to back-pressure.

In summary, the results demonstrate that \tool can effectively co-serve online and offline inference and adapt to different SLOs. It maintains near-optimal P99 online latency while achieving high GPU utilization for offline throughput, closely matching the optimal baseline. Compared to other baselines, on average, \tool reduces P99 online TTFT by 2.37$\times$ and TBT by 2.72$\times$, respectively, while simultaneously increasing offline throughput by 3.00$\times$.

\mysubsection{End-to-End Serving with Real-World Workloads}
\label{sec:eval:real-overall}

\begin{figure}[t]
    \centering
    \includegraphics[width=0.99\linewidth]{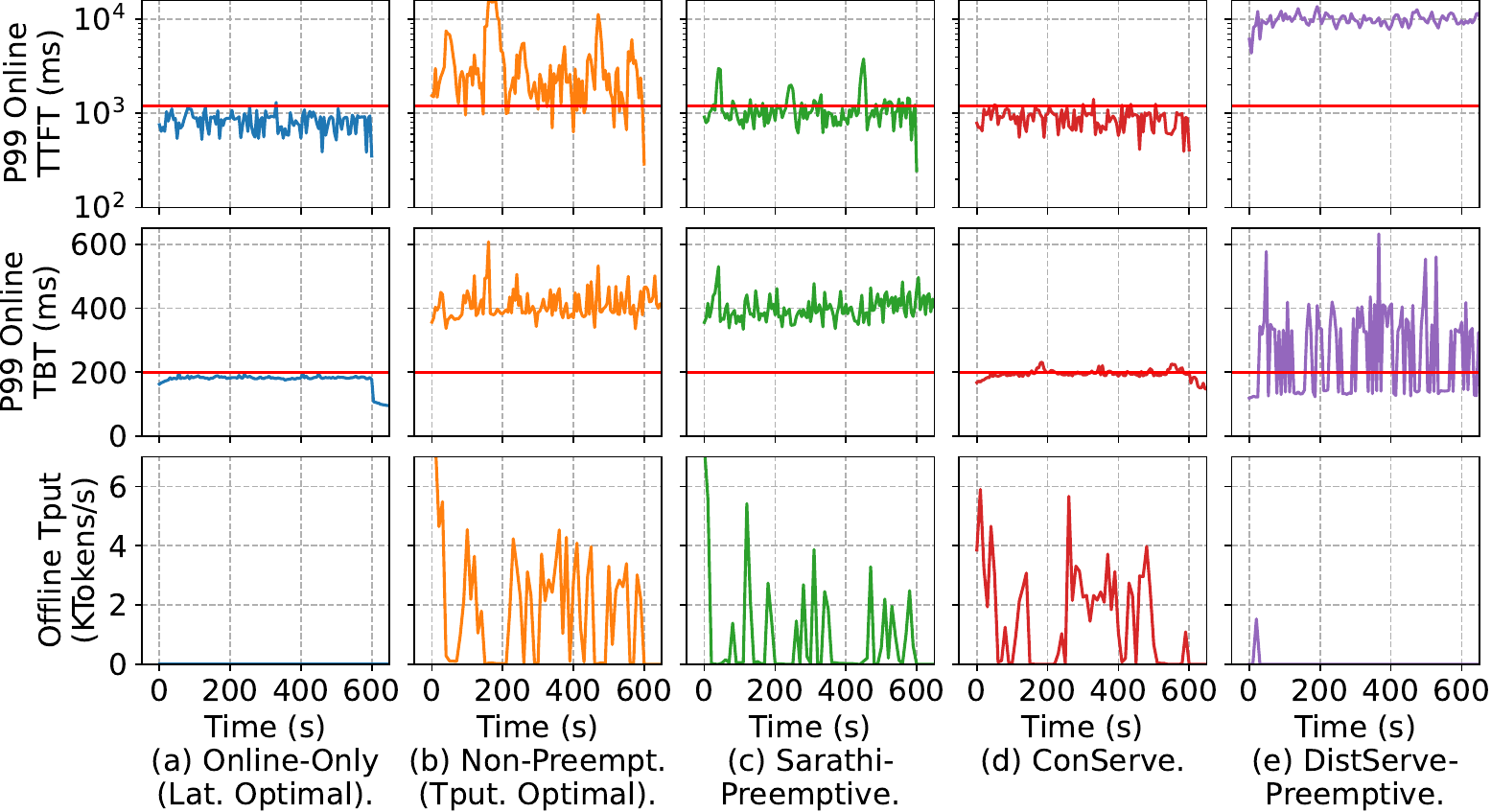}
    \small
    \caption{Serving performance for Llama-3.1 70B on real workloads. \tool is the online system that consistently meets online SLOs (red lines) and achieves 78\% of the ideal offline serving throughput
    (measured with \codeIn{Non-Preemptive}).
    }
    \label{fig:eval:realload-perf}
\end{figure}

\begin{figure}[t]
    \centering
    \includegraphics[width=0.99\linewidth]{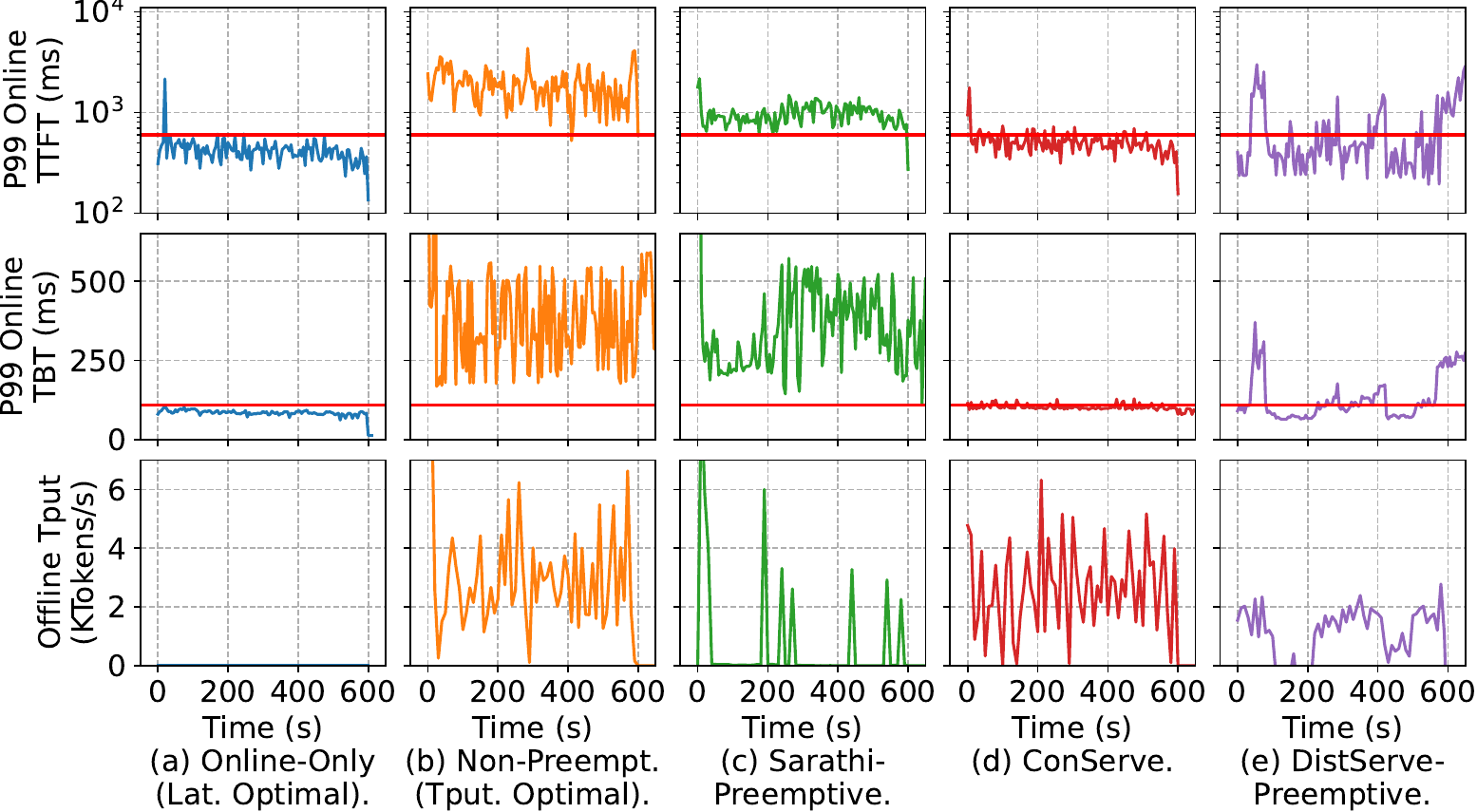}
    \small
    \caption{Serving performance for Llama-3.1 8B on real workloads. \tool maintains its significant lead, achieving 88\% of optimal offline throughput while keeping online latencies low. Although the less-saturated 8B setup is more favorable to \codeIn{DistServe-P}'s architecture, \tool still outperforms it by 2.5$\times$ in throughput and over 3$\times$ in latency.}
    \label{fig:eval:realload-perf-8b}
\end{figure}

This subsection evaluates \tool with real-world workloads and compares its end-to-end performance against the baselines.
We sampled a 10-minute trace from BurstGPT~\cite{burstgpt} as the online workload.
For the offline workload, we use both a random mixture of three distinct offline datasets 
and evaluate co-serving with each dataset individually.

Figure~\ref{fig:eval:realload-perf} presents the time-series results for Llama-3.1 70B. The top two rows present P99 online latencies per 5 seconds, with red lines indicating the respective TTFT and TBT SLOs. 
Both \codeIn{Online-Only} and \tool are able to achieve constantly low P99 TTFT and TBT and meet SLOs. 
In contrast, all three baselines exhibit highly unstable latency and frequently violate TTFT and TBT SLOs.

Regarding offline throughput, \tool is able to achieve an average of 1553 tokens/s, which is 78\% of optimal throughput measured with \codeIn{Non-Preemptive}. Because real-world offline datasets typically feature long sequences, \codeIn{Sarathi-P} triggers frequent, costly recomputations, resulting in a low throughput of only 895 tokens/s.
\codeIn{DistServe-P}'s performance is more nuanced. Under the heavy load of this workload, it performs the worst, as the long sequences exacerbate the back-pressure and preemption issues detailed in \S\ref{sec:eval:synthetic-overall}. However, in less saturated setups, such as with the smaller Llama-3.1 8B model (Figure~\ref{fig:eval:realload-perf-8b}), its architectural imbalances are less pronounced, allowing it to slightly outperform \codeIn{Sarathi-P}. Even in this more favorable scenario, \codeIn{DistServe-P} remains significantly worse than \tool, suffering from 2.5$\times$ lower throughput and over 3$\times$ higher online latency.

Finally, Table~\ref{tab:real-load-overall-results} summarizes results across all offline datasets. Compared to the best-performing baseline, \tool reduces P99 online TTFT and TBT by an average of 2.04$\times$ and 2.86$\times$, respectively, while improving the offline throughput by 2.23$\times$ on average.

In summary, under realistic and dynamic workload conditions, \tool can maintain strict online SLOs while delivering close to optimal offline throughput.

\begin{table}[t]
    \centering
    \begin{adjustbox}{max width=\linewidth}
    \aboverulesep=0ex
    \belowrulesep=0ex
    \begin{tabular}{l|rrrrr|rrrrr|rrrrr}
    \toprule\rule{0pt}{1.1EM}%
    \textbf{Offline} & \multicolumn{5}{c|}{\textbf{P99 Online TTFT (s)}} & \multicolumn{5}{c|}{\textbf{P99 Online TBT (ms)}} & \multicolumn{5}{c}{\textbf{Offline Tput (tok/s)}} \\
    \textbf{Workload} & 
    \codeIn{OO} & \codeIn{NP} & \codeIn{SP} & \codeIn{DP} & Ours & 
    \codeIn{OO} & \codeIn{NP} & \codeIn{SP} & \codeIn{DP} & Ours & 
    \codeIn{OO} & \codeIn{NP} & \codeIn{SP} & \codeIn{DP} & Ours \\
    \midrule\rule{0pt}{1.1EM}%
    Mixed & \multirow{4}{*}{1.08} & 15.2 & 2.91 & 12.4 & 1.14  & \multirow{4}{*}{184} & 450 & 441 & 1398 & 201 & \multirow{4}{*}{0} & 1995 & 895 & 26 & 1553 \\
    DuReader &  & 38.7 & 2.98 & 14.9 & 1.67 &  & 659 & 400 & 1607 & 204 &  & 1276 & 715 & 15 & 1187 \\
    MultiNews &  & 11.0 & 2.11 & 11.3 & 1.04 &  & 305 & 212 & 1384 & 194 &  & 2765 & 1698 & 39 & 2281 \\
    VCSUM &  & 46.2 & 3.01 & 15.1 & 1.68 &  & 566 & 453 & 1774 & 208 &  & 1218 & 255 & 15 & 1064 \\
    \bottomrule
    \end{tabular}
    \end{adjustbox}
    \caption{
    Overall latency and throughput of Llama-3.1 70B when co-serving real-world online and different offline workloads achieved by \codeIn{Online-Only} / \codeIn{Non-Preemptive} / \codeIn{Sarathi-Preemptive} / \codeIn{DistServe-Preemptive} / \tool.
    \label{tab:real-load-overall-results}}
\end{table}

\mysubsection{Performance Sensitivity Analysis}
\label{sec:eval:load-varation}

In this section, we evaluate \tool's robustness to varying workload characteristics, specifically load burstiness and request lengths. We conduct these experiments based on the setup used for Figure~\ref{fig:eval:synthetic-overall-perf}(a), varying one parameter at a time (burstiness, input length, or output length) while keeping other conditions constant.

\begin{figure}[t]
    \centering
    \includegraphics[width=0.99\linewidth]{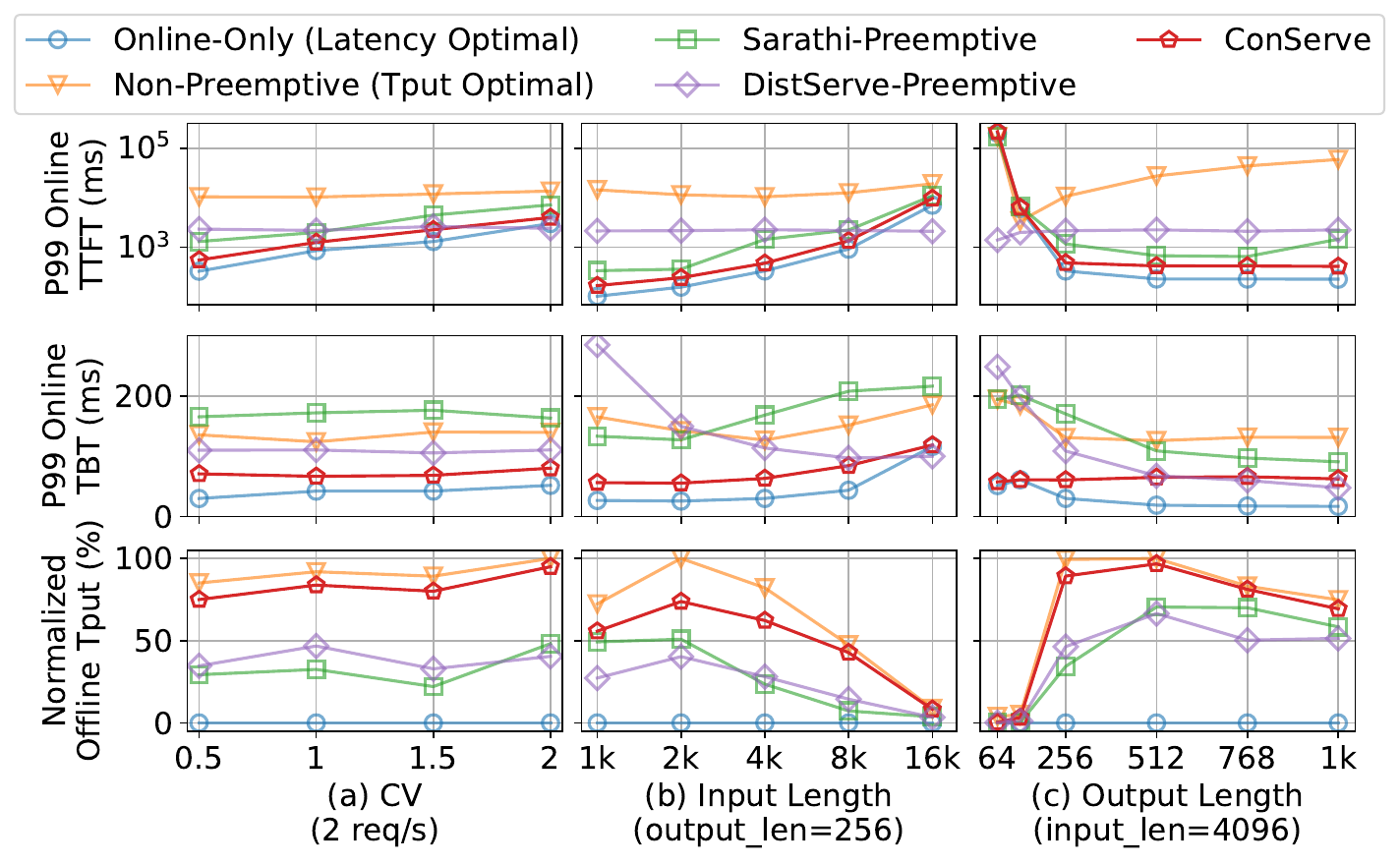}
    \small
    \caption{Llama-3.1 8B serving performance under workloads with varying CVs and input and output lengths.}
    \label{fig:eval:cv-length-overall}
\end{figure}

\MyPara{Load burstiness vs. performance.} We use CV to represent load burstiness, where higher CV indicates greater load variability. As Figure~\ref{fig:eval:cv-length-overall}(a) shows, \tool demonstrates strong resilience to bursty loads with low online latency and high offline throughput that are close to ideal values. Compared to the best-performing baseline, \tool reduces P99 TTFT and TBT by an average of 1.68$\times$ and 2.38$\times$, respectively, and improves offline throughput by 3.62$\times$.

\MyPara{Input/output lengths vs. performance.}
To maintain comparable load when varying sequence lengths, we adjust the request rate inversely proportional to length relative to the \S\ref{sec:eval:synthetic-overall} setup (4096 input tokens, 256 output tokens, at 2 req/s). 

Figure~\ref{fig:eval:cv-length-overall}(b) and (c) show results for varying input and output lengths. \tool delivers good online latency that is close to \codeIn{Online-Only} across most settings. 
Regarding offline throughput, \tool consistently outperforms two preemptive baselines. This advantage is most pronounced for typical requests with input lengths between 2K and 16K tokens, and output lengths longer than 128 tokens. 
We analyze the performance at the extremes.
(1) Requests with short inputs (<2K tokens) cannot fully utilize KV cache capacity. This allows preemptive baselines to keep paused offline requests in GPU memory and avoid recomputation; (2) Requests with long inputs (>16K tokens) or short outputs (<128 tokens) are compute-intensive, leaving little idle capacity for offline serving.
(3) Requests with long outputs (>512 tokens) are less compute-intensive but consume more KV cache because they stay on GPUs longer. This memory pressure limits the number of offline requests that can be served and slightly hinders offline throughput.

In summary, \tool can achieve robust performance across varying degrees of burstiness and request lengths, and it delivers consistently higher offline throughput than baselines, especially for typical-length requests that closely align with real-world workloads.

\mysubsection{Design Drill-Down}
\label{sec:eval:drilldown}

\MyPara{Ablation study.} We quantified the individual contribution of each of our three core optimizations by incrementally enabling them atop the \codeIn{Sarathi-P} baseline, using the Llama-3.1 8B setup from Figure~\ref{fig:eval:synthetic-overall-perf}(a). Compared to \codeIn{Sarathi-P}, our SLO-aware scheduler reduces P99 online TBT from 375ms to 73ms. This shorter iteration time also makes the scheduler more responsive to new requests and reduces P99 online TTFT from 1399ms to 686ms. Enabling layer-wise preemption eliminates TTFT jitter and further reduces P99 TTFT from 686ms to 523ms and worst-case TTFT by 34\% from 866ms to 573ms. Finally, adding incremental KV cache management further improves offline throughput by 1.29$\times$ from 8810 tokens/s to 11366 tokens/s.

\MyPara{Cost of incremental KV cache management.}
We measured the overhead and PCIe utilization of our incremental KV cache mechanism. For Llama-3.1 8B on a single H100 GPU, our kernel can transfer KV caches at 37GB/s, and the largest transfer finishes within 7ms and can be completed overlapped. The kernel introduces a <500\us cost to gather KV caches of scattered tokens, which is negligible compared to model execution. It also scales well in a distributed setting. For Llama-3.1 70B on 4 H100s, each TP worker checkpoints its local KV cache shard concurrently, achieving an aggregated bandwidth of 132GB/s. 

\MyPara{Efficiency of layer-wise preemption.}
\label{sec:eval:drilldown:preemption-cost}
We measured the runtime cost of the preemption safepoint and compared it to the model execution time. On a single GPU, the instrumented safepoint incurs negligible cost (<10\us). Even on four GPUs with TP, each safepoint takes only 167.2\us on average. This is slightly higher than raw instruction cost due to the minor lag among TP workers, but still three orders of magnitude faster than model execution, which takes 262ms on average. 
\tool instruments the model every 4 layers and introduces a total latency increase of 3.0ms when no preemption occurs, accounting for a 1.1\% overhead compared to the model execution time. Meanwhile, \tool maintains responsiveness by detecting new requests and preempting the running batch within 13ms, ensuring timely responses to online load bursts.

\mysection{Related Work}

\MyPara{LLM Serving Systems.}
Prior LLM serving systems mainly focus on optimizing throughput and latency for a single type of workload. To improve throughput, Orca~\cite{orca@osdi22} proposed continuous batching, and vLLM~\cite{vllm@sosp23} proposed PagedAttention. They addressed inefficiencies in scheduling and KV cache management. To reduce latency, Sarathi~\cite{sarathi-serve@arxiv24} and splitfuse~\cite{holmes2024deepspeedfastgen} proposed chunked-prefill to split long prefills into small chunks for better TBT. \tool is built upon these techniques and uniquely co-serves online and offline tasks for maximized GPU utilization. 
DistServe~\cite{distserve@arxiv24} and Splitwise~\cite{patel2024splitwise} disaggregate prefill and decode computation to model replicas for best TTFT and TBT, but they suffer from low GPU utilization because of imbalanced computation load between prefill and decode instances.

\MyPara{Generic Model Serving Systems.}
Before the rise of LLMs, many systems
target serving general ML models~\cite{triton-inference-server@web,torchserve@web,tensorflow-serving@web,clipper@nsdi17,inferline@socc20,clockwork@osdi20,reef@osdi22,alpaserve@osdi23,pipeswitch@osdi20}. 
Among them, Clipper~\cite{clipper@nsdi17} and Clockwork~\cite{clockwork@osdi20} serve general neural networks by batching and scheduling requests. \textsc{Reef}~\cite{reef@osdi22}, \textsc{Shepherd}~\cite{shepherd@nsdi23}, and PipeSwitch~\cite{pipeswitch@osdi20} multiplex a GPU for serving multiple small DNN models using preemptive scheduling. AlpaServe~\cite{alpaserve@osdi23}, on the other hand, leverages model parallelism for statistical multiplexing. However, these systems overlook the huge model size and the autoregressive nature of LLM inference, hence do not support or only achieve suboptimal performance for LLM serving.

\MyPara{Offline LLM Serving.} As offline inference has gained increasing traction, many systems are specifically optimized for offline LLM serving. DeepSpeed ZeRO-Inference~\cite{aminabadi2022deepspeed-inference} and FlexGen~\cite{flexgen@icml23} offload model weights and KV caches to host memory to serve LLMs on small commodity GPUs. 
However, due to the high latency of model swapping, these systems are ill-suited for online serving.
BlendServe~\cite{zhao2024blendserve} and PSM~\cite{liu2024optimizing} improve KV cache reuse and compute efficiency by reordering offline requests within a batch.
\tool is compatible with them for further performance improvements.

\MyPara{Workload Colocation.}
Colocating latency-critical applications with batch applications is a widely adopted approach to improve resource utilization in datacenters.
Various CPU schedulers~\cite{parties@asplos19,shenango@nsdi19,caladan@osdi20} and operating systems~\cite{zygos@sosp17,junction@nsdi24} have been proposed to partition or reassign CPU cores between latency-critical jobs and batch jobs for higher CPU utilization.
Compared to them, \tool is the first to apply the principle of workload co-location specifically for LLM serving.

\mysection{Conclusion}
We present \tool, a new LLM co-serving system that harvests idle GPU resources in online inference for offline serving.
It maintains low online serving latency with an SLO-aware scheduler and a near real-time, layer-wise preemption mechanism. It also achieves high offline serving throughput by incremental KV cache management to avoid recomputation.
Our evaluation shows that \tool significantly outperforms existing state-of-the-art systems.

\bibliographystyle{abbrv}
\bibliography{paper}

\end{document}